\documentclass[twocolumn,twoside,slac_two]{revtex4}
\usepackage{graphicx}
\usepackage{fancyhdr}
\begin{document}

\title{Theory Summary {\large (a Perspective}}

%

\author{George W.-S. Hou}
\affiliation{
Department of Physics, National Taiwan University, Taipei, Taiwan 10617}
%

\begin{abstract}

This is the Theory Summary of the ``Flavor Physics and CP Violation 2012" conference,
with emphasis on New Physics.
Besides covering the theory part of the conference,
we pay attention also to the physics highlights of experimental talks.
I then give my perspective on the false ``Godot sightings" of the past decade,
with some firsthand accounts.
With all coming to naught (well, SM) at the moment,
I look ahead to the near future, and to 2015 and beyond.
An Epilogue is added with the advent of ``the Higgs" at the LHC.
\end{abstract}

\maketitle

\thispagestyle{fancy}


\section{TOWARDS ``PARADIGM SHIFT"}

Once upon a simpler time, ``\emph{la raison d'\^{e}tre} of the B factories"~\cite{Trabelsi}
was very clear: measure\footnote{
 Curiously, though affiliated with KEK and
 a former Belle analysis coordinator,
 Karim Trabelsi did not use the notation of $\sin2\phi_1$ in his talk.}
 $\sin 2\beta$.
Sure enough, just two years' running of Belle and BaBar settled the case:
the measured $\sin 2\beta/\phi_1$ value established the CKM paradigm,
and Kobayashi and Maskawa were awarded the 2008 Nobel prize~\cite{Nobel2008}.
In contrast, even if ``the Higgs" may have emerged at the LHC,
in terms of New Physics (NP), we are still \emph{Waiting for Godot}.

Guy Wilkinson brought \emph{Waiting for Godot} into FPCP 2009~\cite{Wilkinson}
as a parable for the search of New Physics in the flavour sector.
This famous existential play by Samuel Beckett has two main characters.
Vladimir seems to search for meaning and purpose,
while Estragon exemplifies the ignorance of man.
With experimentalists perhaps in protest, I will play Vladimir,
the theorist. However, I will also comment on experiment.

At FPCP 2011, the mood was expectant:
``We are not waiting for Godot anymore ...
with the excellent initial performance of the LHC detectors,
\textbf{we are on our way to find Godot}", exclaimed Frederic Teubert
in his experimental summary.
Guido Altarelli gave a theorist conclusion~\cite{Altarelli},
expressing some anxiety:
``We really hope (the LHC) will start a new era:
not just indirect hints of NP,
but direct production of new states."
However, giving the ``sage" talk two months later,
at the Joint ECFA-EPS Session of the EPS-HEP meeting in Grenoble,
Altarelli sounded shaken:
``Not a single significant hint of new physics found".
This once again echoes \emph{Waiting for Godot},
expressing a sense of \emph{d\'{e}j\`{a} vu},
as Vladimir and Estragon have been \emph{Waiting for Godot}
for God knows how long ...

As much as the CKM paradigm is established, we dream of
a paradigm shift that may arise some day from flavor physics.

\section{OUR NORMAL SCIENCE: A SUMMARY}

Let me begin my theory summary.

One robust theme is spectroscopy, even though it was
not much represented in this conference.
The central theme, of course, is CKM, where the aim is,
by doing it really well, we might get hints towards New Physics.
Lattice has become a great help, while as if through a (distorting)
mirror, we have the ``PMNS" paradigm of the neutrino world.
The theorist needs to study specific modes and processes,
which are pursued by various experiments,
while there is also the direct search agenda for New Physics.
To carry the program further, we need to project towards
the planned new facilities or detector upgrades.

\subsection{Spectroscopy}

The Onia saga, relaunched with the X(3872) discovery in 2003,
has long since turned into an XYZ zoo, and it is still not quite understood.
The fact that both ATLAS and CMS have reported~\cite{Chisholm} new states,
the $\chi_b(3P)$ and $\Xi_b^*$, respectively, reflects the
prowess of the LHC collider experiments, and robust state of the subfield.

For lattice progress on onia, I quote Sinead Ryan directly:
``Charm is more mature than bottom. Watch the next 5 years."
See also the separate summary by Aida El-Khadra~\cite{El-Khadra}.

What may be a little surprising, as covered by Jian-Xiong Wang~\cite{JXWang},
is the hyperactive subfield of onia production.
This is due to the abundance of data, as presented by James Russ~\cite{Russ},
giving theory ``continued guidance".
Just as a sample, the number of PRLs in the past few years by
the groups of K.-T. Chao, J.-X. Wang, and B. Kniel, are impressive,
covering the topics of double charmonia production,
${\cal O}(\alpha_s^4v^4)$ effects, $J/\psi$ polarization, even for the LHC.

\subsection{Non-CKM PV and Rogue Neutrinos}

Can there be $P$ and $CP$ violation in heavy ion collisions?
Kenji Fukushima presented a fascinating talk~\cite{Fukushima} on Local Parity Violation (LPV),
related to the strong $CP$ problem (hence non-CKM sourced).
A chiral magnet effect could lead to charge separation fluctuations
due to the strongest \textbf{B} (QCD scale!) field in the Universe
that is produced by the heavy ion collision.
However, the question is how to experimentally
separate the effect from the observed ``flow".

Neutrinos have been in the news.
Andrew Cohen discussed~\cite{Cohen} superluminous travel of neutrinos,
pointing out that such neutrinos could lead to vacuum \v{C}erenkov
radiation of $e^+e^-$ pairs, but the deviation from speed of light,
as originally claimed by OPERA, is way too large.
However, ``the Fat Lady
has already sung."\footnote{
 For the benefit of local Chinese:
 ``It ain't over 'till the Fat Lady Sings"
 is a colloquial remark originating from American sports,
 referring to the game but alluding to opera endings.
 We note that, seemingly there is never a Fat Lady in Hui Opera,
 as evidenced from the cultural event of the conference.}
It is over: bad cable attachment.
There is, however, some silver lining.
The observation of cosmic neutrinos at hundred TeV scale
leads to very strong new constraints on neutrino Lorentz violation.

We return to CKM and PMNS matrices later.

\subsection{Semileptonic and Leptonic Decays}

Giulia Ricciardi covered semileptonic $B$ and $D$ meson decays~\cite{Ricciardi}.
But she was preceded by three experimental talks, so she commented
that the experimental talks already covered a lot of theory.
Indeed, the field is mature.

The main theme is the long standing tension, typically more than 2$\sigma$,
between exclusive vs inclusive measurements of
\emph{both} $|V_{cb}|$ and $|V_{ub}|$,
as presented very well also by Vera L\"uth's~\cite{Luth} experimental talk
on semileptonic $B$ decays.
The problem of $|V_{ub}|$ is further aggravated by the
large experimental\footnote{
 At ICHEP 2012, Belle reported an update on $B\to \tau\nu$ that
 agrees well with lattice expectations.
 But, albeit somewhat diminished, the tension for $|V_{ub}|$ lingers.}
value for $B\to \tau\nu$, implying an even larger $|V_{ub}|$
than suggested by inclusive data.
Once again, lattice QCD provides valuable, strong input on
form factors and decay constants~\cite{El-Khadra}.
Which value for $|V_{ub}|$ should one take?
This would affect the NP scenario, which we will return to later.

Of course, as covered by Koji Hara~\cite{Hara} from the experimental perspective,
enhanced $B\to \tau\nu$ itself could suggest NP,
specifically an $H^+$ boson from type II 2HDM
(two Higgs Doublet Model), which naturally arises from minimal SUSY.
With charged Higgs $H^+$ possibly mediating the decay
in addition to the $W^+$ boson,
the SM rate is modified by a simple multiplicative factor~\cite{Hou93},
\begin{eqnarray}
r_H = \left[1 - \tan^2\beta \frac{m_B^2}{m_{H^+}^2}\right]^2,
\end{eqnarray}
without further hadronic uncertainties. 
Supersymmetry (SUSY) modifies~\cite{AR03} this mildly,
\begin{eqnarray}
r_H = \left[1 - \frac{\tan^2\beta}{1 + \varepsilon_0} \frac{m_B^2}{m_{H^+}^2}\right]^2.
\end{eqnarray}

But let's turn to recent experimental development.

\vskip0.3cm
\noindent\underline{\bf\boldmath The $B\to D^{(*)}\tau\nu$ Bombshell from BaBar}
\vskip0.2cm

Masked by the ``Semileptonic $B$ Decays" title and the matter-of-fact tone,
L\"uth's talk~\cite{Luth} unleashed a shocker that is for sure
a conference highlight:
a new BaBar result that could shake, indirectly,
the foundations of supersymmetry.
It turned out that BaBar submitted the original paper~\cite{BaBarDtaunu}
to PRL more or less at the same time of the talk, so it is
the first time this potentially important result was reported!
What is remarkable is that this BaBar study even corrected some mistake(s)
by theorists on the subject of $B\to D^{(*)}\tau\nu$,
so it does contain theoretical elements that warrants mention
in this Theory Summary.

BaBar measured the ratios~\cite{Luth,BaBarDtaunu}
\begin{eqnarray}
 {\cal R}(D)
  &=& \frac{\Gamma(\bar B \to D\tau\nu)}{\Gamma(\bar B \to D\ell\nu)} = 0.440 \pm 0.071, \\
 {\cal R}(D^*)
  &=& \frac{\Gamma(\bar B \to D^*\tau\nu)}{\Gamma(\bar B \to D^*\ell\nu)} = 0.332 \pm 0.029,
\end{eqnarray}
where $\ell = e$, $\mu$, and several experimental and theoretical uncertainties cancel.
I will spare experimental details, but both these measured
values 
are higher than SM expectations~\cite{Luth,BaBarDtaunu},
\begin{eqnarray}
 {\cal R}(D)_{\rm SM} &=& 0.297 \pm 0.017,\\
 {\cal R}(D^*)_{\rm SM} &=& 0.252 \pm 0.003,
\end{eqnarray}
with a combined significance of 3.4$\sigma$.

\begin{figure}[t!]
\includegraphics[width=2.8in]{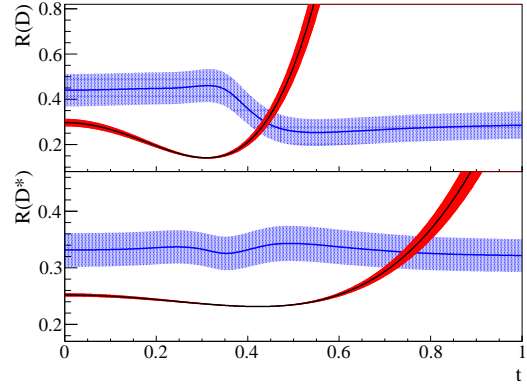}
\caption{
 Comparison of BaBar's $B\to D^{(*)}\tau\nu$ results
 with $H^+$ effect from type II 2HDM
  (the narrow band).
 The $x$-axis is $\tan\beta/m_{H^+}$ in GeV$^{-1}$ units,
with SM at $\tan\beta/m_{H^+} = 0$
 [plot from Ref.~\cite{BaBarDtaunu}].
}
\end{figure}

With good experimental precision, BaBar checked against
the possible interpretation with type II 2HDM.
The differential decay rate formula is~\cite{Luth,BtoDtaunu},
\begin{eqnarray}
 \frac{d\Gamma_\tau}{dq^2}
  &=& \frac{G_F^2|V_{cb}|^2|{\rm\bf p}|q^2}{96\pi^3m_B^2}
      \biggl[1 - \frac{m_\tau^2}{q^2}\biggr]^2
      \biggl(\left[|H_{\scriptsize ++}|^2 + |H_{\scriptsize  --}|^2\right. \nonumber \\
  && \left. + |H_{00}|^2\right]
      \biggl[1 + \frac{m_\tau^2}{2q^2}\biggr]
   + \frac32\frac{m_\tau^2}{q^2}|H_{0t}|^2 \biggr),
\end{eqnarray}
where $q^2$ is $\tau\nu$ lepton-pair mass, $|{\rm\bf p}|$ a momentum
defined in Ref.~\cite{BtoDtaunu} for defining lepton-pair helicities,
$H_{mn}$ are helicity amplitudes with
$D^*$ and lepton-pair helicities $+$, $-$ and $0$,
plus a 4th component $t$ for the latter;
for $B\to D\tau\nu$, $H_{\pm\pm}$ is absent.
The charged Higgs $H^+$ effect enters only through
the last term of Eq.~(7), via
\begin{equation}
 H_{0t}^{\rm 2HDM} = H_{0t}^{\rm SM}
                     \left[1 - \frac{m_b\tan^2\beta}{m_b \mp m_c}
                               \frac{q^2}{m_{H^+}^2} \right],
\end{equation}
where $-$ ($+$) sign is for $B \to D^{(*)}\tau\nu$.
Compared with Eq.~(1),
the $m_b/(m_b \mp m_c)$ factor brings in hadronic uncertainties,
albeit not too severely.
We note that the numerator and denominator are of different origins.
Note also that $H_{0t}^{\rm SM}$ contains the scalar form factor that
does not appear for $\bar B \to D\ell\nu$.

Accounting for difference in efficiency for ${\cal R}(D)$ and  ${\cal R}(D^*)$
measurement for twenty different $\tan\beta/m_{H^+}$ values,
the BaBar result is plotted in Fig.~1 vs $\tan\beta/m_{H^+}$,
compared with the expected theoretical values.
The two intersections are~\cite{Luth,BaBarDtaunu}
\begin{eqnarray}
 \tan\beta/m_{H^+} &=& 0.44 \pm 0.02\ {\rm GeV}^{-1},\ \, [{\cal R}(D)], \\
 \tan\beta/m_{H^+} &=& 0.75 \pm 0.04\ {\rm GeV}^{-1},\ \, [{\cal R}(D^*)],
\end{eqnarray}
with impressive precision because many uncertainties cancel.
The two numbers are incompatible with each other,
and the combination of ${\cal R}(D)$ and ${\cal R}(D^*)$
``excludes the type II 2HDM charged Higgs boson with a 99.8\% confidence level
for any value of $\tan\beta/m_{H^+}$"~\cite{Luth,BaBarDtaunu}.

This is an astounding statement. \emph{What if the two values met\,!?}
Actually, whether they would meet at the first, or second value,
it would be in very strong conflict with the measured $B\to \tau\nu$ rate:
\emph{the $\tan\beta/m_{H^+}$ values seem too large},
when seen in the light of the $m_B^2$ factor in Eq.~(1).
Within type II 2HDM, if we take $\tan\beta/m_{H^+} \simeq 0.44$ [0.75] GeV$^{-1}$
from Eq.~(9) [(10)], the $B\to \tau\nu$ rate would be enhanced by $r_H \sim 19$ [215].
This is not only ruled out by direct measurement~\cite{Hara},
it would add a challenge to the interpretation of Eqs.~(3) and (4).
It also illustrates that the deviation in ${\cal R}(D^*)$ is even more problematic.
Put another way, given that $B\to \tau\nu$ rate is of order SM expectation,
$\tan\beta/m_{H^+} \lesssim \sqrt2/m_B < 0.27$ GeV$^{-1}$, which is
considerably less than Eq.~(9), and should somewhat suppress
${\cal R}(D)$ and  ${\cal R}(D^*)$ compared with Eqs.~(5) and (6).
$B\to \tau\nu$ is the most sensitive of the three processes
to charged Higgs boson of type II 2HDM,
which was a point emphasized in Ref.~\cite{Hou93}.
I therefore suspect there is a loophole somewhere.

In any case, we await the result from a similar analysis at Belle
(although Belle's measurement of $B \to D\tau\nu$ and $B \to D^{*}\tau\nu$ rates
are also on the large side),
and theorists must check all the assumptions made.
If BaBar's statement pans out, then it would be
a further blow to minimal SUSY, in that the
Higgs sector is more complicated than the minimal type II 2HDM.
With simplicity lost, having more parameters
does not make it more appealing in interpreting the BaBar findings.

\subsection{Kaons --- the Origins of CKM}

We switch to kaons because of the similarity in physics and processes.

The kaon sector is truly the granddaddy of flavor and $CP$ physics
that is this conference. Indeed, much of the CKM structure was
learnt from studying kaon mixing, CPV, and rare decays.
As stressed by Giancarlo D'Ambrosio~\cite{Ambrosio},
it was in facing the kaon system that
the SUSY flavor problem arose, which lead to
the suggestion of MFV (Minimal Flavor Violation),
i.e. all sources of ``FPCP" are rooted in CKM.

\vskip0.3cm
\noindent\underline{\bf\boldmath $K\to \ell\nu$ and Lepton Universality}
\vskip0.2cm

Evgueni Goudzovski discussed~\cite{Goudzovski} the process
$K^+ \to \ell^+\nu$, which is analogous to $B\to \tau\bar\nu$.
With charged Higgs $H^+$ possibly mediating the decay,
one simply replaces $m_B$ in the $r_H$ factors of Eqs.~(1) and (2) by $m_K$.
Given the precision of kaon measurements,
this was refined further for loop effects
involving slepton-sneutrino-bino~\cite{Masiero08},
which motivated the NA62 experiment to measure the ratio
$R_K = \Gamma(K \to e\nu)/\Gamma(K \to \mu\nu)$ and test lepton universality.
With data taken during 2007-2008 in the ``$R_K$ phase" of NA62 running,
the measured value of
$R_K = (2.488 \pm 0.010) \times 10^{-5}$
is in rather good agreement with
$R_K^{\rm SM} = (2.477 \pm 0.001) \times 10^{-5}$~\cite{Cirigliano}.
Depending on the slepton mixing parameter $\Delta_{13}$,
this can rule out extra regions of the $m_{H^+}$--$\tan\beta$ plane,
beyond those from $B\to \tau\nu$ and $b\to s\gamma$.

It should be clear that people are pursuing this type of
searches for $H^+$ effect.
On the other hand, as stressed in previous subsection,
we all now have to contend with the BaBar claim of
ruling out the whole $m_{H^+}$--$\tan\beta$ plane
by their $B\to D^{(*)}\tau\nu$ result.

\begin{figure}[t!]
\includegraphics[width=3.2in]{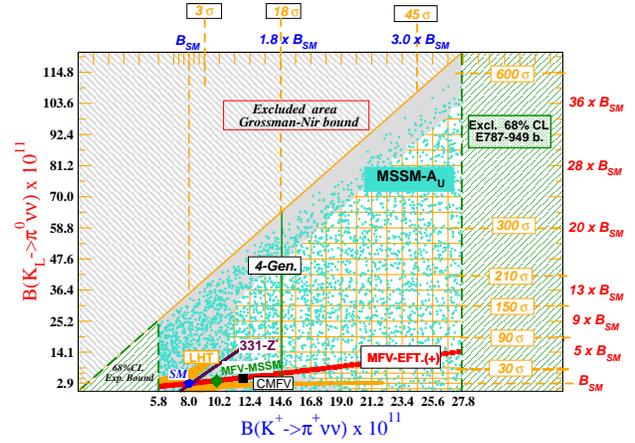}
\caption{
 Mescia-Smith plot~\cite{Krare_BSM} of
 $K^+ \to \pi^+\nu\nu$ vs $K_L \to \pi^0\nu\nu$ physics sensitivities.
}
\end{figure}

\vskip0.3cm
\noindent\underline{\bf\boldmath The Holy Grail: $K\to \pi\nu\nu$}
\vskip0.2cm

As titled, the final goal and quest of kaon physics is to measure $K \to \pi\nu\nu$.
I show in Fig.~2 the ``Mescia-Smith" plot~\cite{Krare_BSM}
to illustrate the merits of
$K^+ \to \pi^+\nu\nu$ vs $K_L \to \pi^0\nu\nu$ measurement.
One sees the bound from the E787/E949 experiment at BNL.
In a way, the aforementioned $R_K$ study by NA62 is
for preparatory purposes. The goal of NA62~\cite{Goudzovski} is to have
${\cal O}(100)$ $K^+ \to \pi^+\nu\nu$ decay events
with $\sim 10\%$ background in 2 years of data taking.
The first technical run is expected in October 2012.

From Fig.~2, however, it should be clear that
$K_L \to \pi^0\nu\nu$, which is a purely CPV decay,
is probably a better probe of NP.
In pursuit of this, the KOTO experiment at J-PARC
aims to reach the Grossman-Nir bound
(allowed by E787/E949 result on $K^+ \to \pi^+\nu\nu$)
with data to be taken, at 10\% intensity, during 2012 Japan Fiscal Year (JFY),
which is most likely during 2013.
With this demonstrated, one would truly be in business to probe NP,
and the plan~\cite{KOTO} is to have extended runs for 2013-2017 JFYs,
to reach eventually down to SM expectations.

\subsection{\boldmath Hadronic $B$ and $B_s$ Decays}

This was a hot subject in first half of the 2000's, during the rising phase
of the B factory era. But, in my view, QCDF turned ``process-dependent"
(allowing hadronic parameters); pQCDF seems under-recognized;
SCET is a pretty fa\c{c}ade to behold, but got the $\Delta {\cal A}_{K\pi}$
(will discuss soon) all wrong, and as far as I know,
never revisited it since FPCP 2008 (even 2007).
I cannot do justice summarizing this vast subject, so let me
just paraphrase Cai-Dian L\"{u}~\cite{CDLu} regarding amplitudes, that
 1) $T$ (color-favored tree): as expected;
 2) $C$ (color-suppressed tree): turned out unexpected
    (i.e. experiments revealed it to theorists,
     who never predicted it);
 3) Annihilations ($A$, $E$, $PA$, $SP$):
    now we know what we did not expect to know ...;
 4) $P$, $P_{\rm EW}$: so many kinds of them!

We will see more of this in the next subsection.
A new development was reported by Yue-Liang Wu
on a 6-quark effective Hamiltonian approach~\cite{YLWu},
by connecting the traditional 4-quark operators to
another quark line via a gluon.\footnote{
 What should one do for non-resonant 3-body decays?
 }
Though different, it seems to be a mixture of
QCDF, pQCDF, SCET and even NF (Naive Factorization).
While it is certainly not easy at all to construct a competing theory,
but like all others, this theory needs to be verified (proof of validity),
and should be checked against the number of assumptions made
versus the number of predictions, and whether the predictions get confirmed.

\subsection{\boldmath D Meson DCPV Difference: $\Delta A_{CP}$}

Michael Gronau accounted~\cite{Gronau} for the frenzied theory activity
in the past 1/2 year on the subject, with equal spread between
SM and NP, oftentimes both.
This frenzy started when LHCb~\cite{Vagnoni,DACP_LHCb} reported nonzero
$\Delta A_{CP} \equiv A_{CP}(K^+K^-) - A_{CP}(\pi^+\pi^-)$ in late 2011 with 0.6~fb$^{-1}$ data,
now followed by CDF's recent update~\cite{DACP_CDF} to 9.7~fb$^{-1}$ data,
namely,
\begin{eqnarray}
 \Delta A_{CP} &=& (-0.82 \pm 0.21 \pm 0.11)\%,\ \ {\rm (LHCb)} \\
 \Delta A_{CP} &=& (-0.62 \pm 0.21 \pm 0.10)\%,\ \ {\rm (CDF)},
\end{eqnarray}
at 3.5$\sigma$ and 2.7$\sigma$ respectively,
as reported by Vincenzo Vagnoni~\cite{Vagnoni}.
I congratulate CDF for being competitive!

\vskip0.3cm
\noindent\underline{\bf SM or NP?}
\vskip0.2cm

Although the before-the-fact anticipation was tiny,
the statement from Gronau, not unexpectedly, is that
$\Delta A_{CP}$ is \emph{not inconsistent} with SM, but it needs
the $c\bar u \to u\bar u$ penguin (which pops out a $\bar ss$ and $\bar dd$ pair
to make the $K^+K^-$ and $\pi^+\pi^-$ final state) to be enhanced by
$\sim 10$ compared to the naive estimate.
By first tuning the $T$ amplitudes to Cabibbo allowed decays,
it is found that experimentally measured $D^0 \to K^+K^-$
vs $D^0 \to \pi^+\pi^-$ ratios indicate large $U$-spin (a subset of SU$_{\rm F}$(3))
breaking~\cite{Gronau12} in singly-Cabibbo-suppressed (SCS) tree decay,
the leading process.
These authors then suggest that $U$-spin breaking in $P + PA$
(the latter is ``Penguin Annihilation",which we have seen in hadronic $B$ decays)
could lead to the needed enhancement from the naive tiny value.
I refer to Gronau's talk for further discussion.

In the discussions after the talk, Hai-Yang Cheng asked, ``What about SU(3) breaking in $E$?"
Note that $E$ is a form of annihilation with $W$ boson exchange that converts $c\bar u \to d\bar d$
(with $\bar ss$ and $\bar dd$ popping from vacuum),
and always come together with the SCS $T$ amplitude, sharing the same CKM factor.
In response, Gronau replied ``... simplifying assumptions ...",
which illustrates the somewhat cherry-picking nature
on one's choice of set of amplitudes to work with (or break).

Here we have \emph{d\'{e}j\`{a} vu} again.
Once upon a time, there was (and still is) the
$\Delta {\cal A}_{K\pi}$ (or $K\pi$) puzzle~\cite{BelleNature,Trabelsi, Wilkinson},
the direct CPV (DCPV) difference observed between
$B^+\to K^+\pi^0$ and $B^0\to K^+\pi^-$.
It was suggested in similar fashion that this could be
due to enhanced $C$, which, being the color-suppressed tree,
it was naively expected to be tiny beforehand.
However, with no indication of NP in $B_s$ TCPV (time-dependent CPV,
from mixing-decay interference) so far, as reported by Yuehong Xie~\cite{Xie},
perhaps enhanced $C$ is behind $\Delta {\cal A}_{K\pi}$,
without the need for NP.

With the $D$ meson system much more susceptible to ``hadronic effects" than the $B$ system,
we expect many more theory papers on $\Delta A_{CP}$ to come,
but it would be hard to settle which approach is correct.
Because of this, even if it would appear more and more like a NP source after further scrutiny,
the hadronic mess would unlikely allow one to point back
to indentify ``\emph{What NP?}".
However, predictions of model approaches should be
followed, which would certainly stimulate a lot of measurements.
And that is certainly the standard form of our ``normal science".

\begin{figure}[t!]
\includegraphics[width=2.9in]{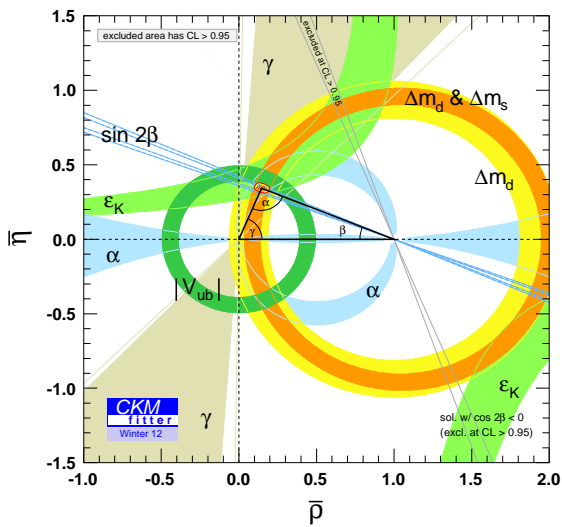}
\caption{
 Global CKM fit up to and including Moriond 2012 results.
}
\end{figure}

\subsection{CKM Paradigm and PMNS ``Mirror"}

The CKM matrix is the meeting point of experiment and theory,
as reflected in the composition of, e.g. the CKMfitter group.
The neutrino sector ``mirror", the PMNS matrix,
now probably should consider seriously a ``fitter approach",
i.e. taking unitarity into full account, given a third large mixing angle.

\vskip0.3cm
\noindent\underline{\bf CKM Fit: Whither Tensions?}
\vskip0.2cm

As reported by S\'{e}bastien Descotes-Genon~\cite{Descotes-Genon},
fitting all available data to extract info on CKM parameters is now
a regular way to spot ``tension", hence uncover or constrain NP.
The current global fit, incorporating Moriond 2012 results, is given in Fig.~3.

Among the myriad of issues touched upon by Descotes-Genon,
the main themes are the tensions of
\begin{itemize}
 \item $\sin2\beta$ vs $B\to \tau\nu$
(and as can be seen from Fig.~3, a tension between $\varepsilon_K$ and $|V_{ub}|$),
where, besides experimental error or lattice measurement correlations,
could be new physics in $B$ decay, or in mixing;
 \item $A_{SL}$: D$\emptyset$ vs SM expectation;
 \item $(\beta_s,\ \Delta\Gamma_s)$: LHCb measurements now consistent
       with SM, which spotlights $A_{SL}$ of D$\emptyset$.
\end{itemize}
I will return to these issues in my own ``Perspective".

In passing, I mention the point that, because of finite $\Delta\Gamma_s$
as compared to $\Gamma_s$, one needs to be careful in relating
theoretical branching ratios with experimentally measured ones.
For example, for $B_s\to \mu^+\mu^-$, the relation with untagged result is
\begin{eqnarray}
{\cal B}(B_s\to \mu\mu)_{\rm theo} \simeq 0.91 {\cal B}(B_s\to \mu\mu)_{\rm exp,\,untag},
\end{eqnarray}
where 0.91 is roughly $1 - y_s$, with $y_s = \Delta\Gamma_s/2\Gamma_s$.
See the work of De Bruyn \emph{et al.}~\cite{DeBruyn} for details.

\vskip0.3cm
\noindent\underline{\bf The World of PMNS, and MEG}
\vskip0.2cm

Werner Rodejohann~\cite{Rodejohann} presented the status of the PMNS matrix,
the counterpart or mirror of CKM matrix in lepton sector.
It was in this talk that we got to see the explicit
numerical $3\times 3$ CKM mixing matrix, which is amusing.

The big news from earlier this year is the discovery of $\theta_{13} \neq 0$,
which is now above 7$\sigma$ level, with mean value $\sim 8.8^\circ$.
This is in strong contrast to the trickling down nature for CKM matrix
as one goes off-diagonal (cf. $\theta_{13} \sim 0.2^\circ$ for CKM).
When I first heard the result, I exclaimed to my Daya Bay experiment colleague that
``The Postman not only Rings Twice, but Thrice for neutrinos!"

The strength of $\theta_{13} \neq 0$ makes much possible for $\nu$ physics,
but this is not yet ``FPCP core business".
A link with FPCP core business does emerge, as reported by Francesco Renga~\cite{Renga}
on experimental searches for lepton flavor violation (LFV) with charged leptons.
The example of SUSY SU(5) with right-handed neutrinos~\cite{Hisano09} was given,
where the relative rates of LFV in $\mu$ and $\tau$ decays,
such as $\mu\to e\gamma$ and $\tau \to \mu\gamma$, depend strongly on the flavor structure of NP.
With large $\sin\theta_{13}$, hence $U_{e3} \sim \sin\theta_{13} \sim 0.15$,
though $\tau \to \mu\gamma$ is pushed out of reach, $\mu\to e\gamma$ falls into
quite interesting range for the next runs of the MEG experiment,
where 2011 data should reach the $10^{-12}$ level.
The effect is in fact more generic~\cite{Rodejohann}.
For example, with MFV in lepton sector,
${\cal B}(\mu\to e\gamma) \propto |(m_\nu m_\nu^\dag)_{e\mu}|^2$,
and with large $\theta_{13}$, hence finite $U_{e3}$,
$(m_\nu m_\nu^\dag)_{e\mu}$ cannot vanish~\cite{Rodejohann12},
and the decay is guaranteed to occur.
At what strength it does occur is to be probed by experiment.

It would be of high interest to follow the developments in the coming few years.

\subsection{New Ideas and Directions}

The number of theory papers generated by $\theta_{13} \neq 0$
is much larger than $\Delta A_{CP} \neq 0$. Just compare the
respective reference lists in~\cite{Rodejohann} vs~\cite{Gronau}.
Covering ``New Ideas and Directions", Amarjit Soni~\cite{Soni}
stated that
\begin{equation}
\#\ {\rm of\ BSMs}\ \; \sim [\#\ {\rm of\ theorists}]^{{\rm Huge}\#}\ \ !!
\end{equation}
which I certainly agree. But as we emphasized,
determining what phenomenon needs NP is part of the art. Making his pick,
Soni emphasized $\sin2\beta$ vs $V_{ub}$, which I again defer to my Perspective.
Another topic is $\Delta A_{CP}$ and $U$-spin breaking,
which has already been covered by Gronau~\cite{Gronau}, with similar conclusions.
But, in reference to large hadronic (nonperturbative) uncertainties,
for the long term we are reminded of the ``Ghost of $\varepsilon'/\varepsilon$".
The well known and well measured kaon DCPV effect suffered much hadronic uncertainties,
hence falls short from establishing NP.

Not unexpectedly, Soni switched gears to warped extra dimensions (WED, i.e. Randall--Sundrum or RS),
stressing it as an elegant solution to hierarchy and flavor puzzles.
Where he and I resonate is the ``simplest scenario"~\cite{Soni}
of the 4th generation (4G), which may be linked to RS by strong-weak duality.
One very important repercussion from RS is enhanced $t\to cZ$~\cite{Soni,Soni04-06},
\begin{equation}
{\cal B}(t\to cZ) \sim 10^{-5}\left(\frac{\rm 3\ TeV}{m_{KK}}\right)^4
                              \left(\frac{{U_R}_{23}}{0.1}\right)^2,
\end{equation}
where $m_{KK}$ is the KK particle mass scale,
and $U_R$ is an effective right-handed rotation matrix.
The point is that the latter two factors could be $\sim 1$.
However, this seems optimistic now, since we see no sign of NP at the LHC.
In the ``dual" approach of the 4th generation, my old work~\cite{Hou06} shows that
${\cal B}(t\to cZ)$ should be considerably below $10^{-5}$.
Given today's stringent bounds on $m_{b'}$ and the absence of
NP hint in $B_s$ system, $|V_{cb'}^*V_{tb'}| \simeq |V_{t's}^*V_{t'b}|$
should be less than the 0.02 value used in~\cite{Hou06}.
There is certainly experimental interest in $t\to cZ$,
with current best limit of ${\cal B}(t\to cZ) < 0.0034$ from CMS
presented by Vincenzo Chiochia~\cite{Chiochia}.
Scaling from 4.6 fb$^{-1}$ data used for this analysis,
even $10^{-5}$ seems unreachable at the LHC.

Could the enigmatic $A_{\rm FB}^{t\bar t}$ from the Tevatron
be caused by~\cite{Soni} 4G with FCNC scalars?
Well, I would not bet on it.
A general remark on $A_{\rm FB}^{t\bar t}$ would come in the ``Perspective",
to which I now turn.

\section{HOPES \& WISHES --- A PERSPECTIVE}

\emph{Who Moved My Cheese?}
Or, \emph{Where is My New Physics?}
Godot has not yet come, and there is no sure sign of New Physics.
Recalling Eq.~(14), we have certainly not covered all possible NP models and directions.
For these, I refer to the Sage of Flavour,\footnote{
 At the conference, I showed a picture of Andrzej Buras in Bavarian Academician robe.}
who has put forth the ticking Flavour Clock~\cite{Buras12}.

Now, I offer my own hopes and wishes for ``Paradigm" Shift.

\vskip0.3cm
\noindent\underline{\bf Pozzo Comes, Returns Blind}
\vskip0.2cm

I had a couple of personal Godot sightings, which alas,
all turned false.
A true ``It's Godot, we're saved!" moment~\cite{Wilkinson} came,
when Belle saw a 3.5$\sigma$ deviation from expected TCPV in $B^0 \to \phi K_S$
with 140 fb$^{-1}$ data, even of opposite sign w.r.t. $B^0 \to J/\psi K_S$!
This was reported by Tom Browder at Lepton-Photon Symposium at Fermilab in 2003,
and I really held hope that ``This is it."
Well, BaBar did not see such effect at the time,
and, the equal amount of data, taken with the new SVD2
detector at Belle from 2003-2004, gave \emph{opposite sign};
we checked that the probability for this happening is $\sim 4\%$, not small.
The so-called $\Delta {\cal S}$ problem, the deviation between TCPV
as measured in penguin-dominant $b\to s\bar qq$ processes vs $b\to c\bar cs$ processes,
continues to fade to this day.

A second strong indication~\cite{Wilkinson} for NP came from
the aforementioned DCPV difference~\cite{BelleNature}
$\Delta {\cal A}_{K\pi} \equiv {\cal A}_{B^+\to K^+\pi^0} - {\cal A}_{B^0\to K^+\pi^-}$.
Naively, both processes are strong penguin ($P$) dominant. Interference with
the CPV phase carrying tree ($T$) amplitude generates the DCPV, hence
${\cal A}_{B^+\to K^+\pi^0} \sim {\cal A}_{B^0\to K^+\pi^-}$ was expected.
The fact that $\Delta {\cal A} > -{\cal A}_{B^0\to K^+\pi^-} \sim 10\%$,
i.e. ${\cal A}_{B^+\to K^+\pi^0}$ and ${\cal A}_{B^0\to K^+\pi^-}$ seem to differ even in sign,
caused a puzzle~\cite{Trabelsi}.
The culprit could be either the color-suppressed $C$ being enhanced
(and carry a rather different strong phase from $T$),
or there is NP in the subdominant electroweak penguin ($P_{\rm EW}$).
This $\Delta {\cal A}_{K\pi}$ puzzle was marked by Wilkinson~\cite{Wilkinson}
also as a questionable sighting.
Although things could have gone differently since his 2009 summary, he is now likely correct,
given that LHCb sees no indication for NP in $B_s$ TCPV~\cite{Xie},
nor in $A_{\rm FB}(B\to K^{*0}\mu^+\mu^-)$, as reported by Nicola Serra~\cite{Serra}.

I feel particularly sad since both the above ``false Godot sightings" happened
in my group at Belle. Although ${\cal S}_{\phi K_S}$ is a good reminder that most
early indications of NP eventually disappear,
$\Delta {\cal A}_{K\pi} > -{\cal A}_{B^0\to K^+\pi^-}$ is experimentally firm.
Hints of it were already present in 2004, when ${\cal A}_{B^0\to K^+\pi^-}$
was first measured between Belle and BaBar.\footnote{
 Just when ${\cal S}_{\phi K_S}$ measurement suffered a large fluctuation.}
At that time, being shocked, and because of a hunch that
this could be a harbinger of 4th generation $t'$ quark effect through $P_{\rm EW}$,
I embarked on a mission~\cite{Hou0507}.
The point is, by nondecoupling of $t'$ in the loop and bringing with it a new CPV phase,
the 4G effect in $P_{\rm EW}$ could resolve the $\Delta {\cal A}_{K\pi}$ puzzle.\footnote{
 We showed~\cite{HLMN07} further in pQCDF at NLO ($\sim$ QCDF at NNLO) that
 a combined $C$-$P_{\rm EW}$ effect could be at work.}
As important corollaries of nondecoupling,
predictions were made for large ${\cal S}_{\psi\phi}$ ($B_s$ TCPV),
suppressed TCPV in $D$ mixing, and good likelihood that $K_L \to \pi^0\nu\nu$ could be much enhanced.
I therefore went to Fermilab and CERN in Spring 2007 to evangelize,
stressing especially to Tevatron experiments that
there is hope for glory, precisely if the strength of ${\cal S}_{\psi\phi}$
was above 0.5, as might be suggested by $\Delta {\cal A}_{K\pi}$.

Then came 2008, when first CDF, then D$\emptyset$ both reported indications for sizable
${\cal S}_{\psi\phi}$. Before long, the UT\emph{fit} collaboration rushed to state,
 ``It's Godot, we're saved!", or, ``observation of anomalously high CPV in $B_s$ system".
This history was touched upon by quite a few speakers at this conference.
``Not so fast!", said Wilkinson~\cite{Wilkinson} in his FPCP 2009 summary,
when there were no data yet from LHC:
``UT\emph{fit} performed a valuable service to the community by highlighting this intriguing hint,
but combinations are best left to the experiments themselves."
He was again sensible, since the 2010 value of CDF went down a bit.
However, even LHCb's initial result, with only 36 pb$^{-1}$ of 2011 data,
indicated~\cite{Golutvin11} that $-\sin\phi_s$
(equivalent to $\sin2\beta_s$ used by CDF, and what I generically called $-{\cal S}_{\psi\phi}$)
was larger than 0.5.
With much anticipation therefore,
then came what I would call the ``LHCb massacre" at Lepton-Photon 2011:
\emph{All flavor/CPV hints for Godot were killed off!}
Note that this does not affect the possibility of very enhanced $K_L\to \pi^0\nu\nu$.

\vskip0.3cm
\noindent\underline{\bf Some Further Comments}
\vskip0.2cm

We'll still wait for Godot; ``He'll come tomorrow.", as in the play of Beckett.
Here I offer some comments on what Godot may first look like when appearing at a distance:
\begin{itemize}
 \item $A_{\rm FB}^{t\bar t}$: Dubious \\
  The effect suggested by Tevatron data is so large that it must arise from tree level.
  But LHC has pushed NP to above TeV scale.
  The balancing act (for any NP model) seems too tough to me.
  So, what is it that the Tevatron uncovered?
 \item $A_{SL}$: Dubious \\
  The $A_{sl}^b$ as measured by D$\emptyset$ deviates from SM by 3.9$\sigma$,
  as reported by Rick Van Kooten~\cite{VanKooten}.
  When combined with $a_{sl}^d$ as measured at the B factories,
  it suggests a rather large $a_{sl}^s$ value.
  Given that $a_{sl}^s$ is proportional to $\tan\phi_s$ as well as $\Delta\Gamma_s$,
  these two quantities must both\footnote{
   Motivated by the need for enhanced $\Delta\Gamma_s$ above $\Delta\Gamma_s^{\rm SM}$,
   a detailed study~\cite{CHS11} of $b\bar q \to cs\bar c\bar q$ ($q = s$ and $d,\ u$) data
   also found no indication of OPE-violating enhancement.}
  be large, which are now both ruled out by LHCb~\cite{Xie}.
  But, because of the $A_{SL}$ problem, the other possibility of
  New Physics in $B_s$ \emph{decay} ($\Gamma_{12}^s$) was considered seriously
  by Descotes-Genon~\cite{Descotes-Genon} and Soni~\cite{Soni}.
  However, this reminds me of the old ``NP in $D_s \to \mu\nu$" suggestion,
  when compared with $f_{D_s}$ values from lattice.\footnote{
   At FPCP 2008 held in Taipei, I commented
    ``The Lord would be malicious if NP first appears in $D_s \to \mu\nu$."
   at end of Sheldon Stone's talk; ``Thank you, Albert!", he shot back.
   The discrepancy had largely disappeared by FPCP 2010.}
  I would not vouch for it.
 \item LHCb Trio: \underline{$\sin\phi_s$; $B_s \to \mu\mu$; $A_{\rm FB}{(B\to K^{*}\mu\mu)}$} \\
  These are the three premium probes for NP in flavor sector in the LHC era.
  But, while B factories indicated some deviation from SM, as mentioned,
  $A_{\rm FB}{(B\to K^{*}\mu\mu)}$ was the first to conform with SM again.
  We have also mentioned that, while there was high hope for $\sin\phi_s$
  to deviate significantly from zero during 2008-2011, it is now also SM-like,
  and requires high precision (by LHCb) to probe further.
 There has also been rapid progress on $B_s \to \mu\mu$.
 LHCb and CMS quickly ruled out an indication of sizable $B_s \to \mu\mu$ from CDF during summer 2011.
 At Moriond, LHCb and CMS have marched within sight of the SM value,
 as reported by Nicola Serra~\cite{Serra} and Guoming Chen~\cite{GChen}.
 In fact, one might say that there is a mild hint for suppression below SM expectation.
 Given the huge lever arm provided by the $\tan^6\beta$ enhancement with SUSY,
 I would quote a Chinese saying, ``Why kill a chicken with a cow chopper?",
 that it is not quite natural to invoke SUSY for finely tuning ${\cal B}(B_s \to \mu\mu)$
 to below SM values. Instead, we showed~\cite{HKX12} that 4G would provide
 a relatively easy adjustment within the CKM framework.
 We remark that, because of a faster data rate, CMS might overtake LHCb in observing this mode,
 which should be watched.
 \item ``Flavorful SUSY" \\
  With stringent limits placed by the LHC on gluino and light flavored squark production,
  SUSY has been struggling to stay ``Natural".
  One outcome is to have the stops ($\tilde t_1$ and $\tilde t_2$) and the left-hand-$b$ squark light,
  while the gluino and light flavored squarks are heavy.
  This has been dubbed ``Flavorful SUSY"~\cite{PRW11}.
  However, this would break the foundations of MFV,
  briefly discussed by Giancarlo D'Ambrosio~\cite{Ambrosio}.
  So far I do not know of any flavor or CPV predictions coming from ``Flavorful SUSY".
 \item \underline{Higgs as Godot}\\
  The status of Higgs boson search was given~\cite{Charlton}
  by ATLAS Deputy Spokesperson, Dave Charlton.
  It should be emphasized that even 3$\sigma$ is nothing. We have seen so many
  such, or even stronger, indications go away,
  for instance the scary 14y GeV ``Higgs" at EPS-HEP 2011 in Grenoble.
  So, the ``Higgs" hint as of December 2011 at 125 GeV may well be another ``Pozzo"
   (character in Beckett's play on Godot).

  However, Higgs, or No Higgs --- we would know in December for sure ---
  it would impact on FPCP!! [we return for comment in the Epliogue]
\end{itemize}

\vskip0.3cm
\noindent\underline{\bf An Illustration towards 2015+}
\vskip0.2cm

So we have no \emph{serious} sign of Godot in FPCP ...
Besides continuous pursuits of $B_s \to \mu\mu$, $\sin\phi_s$, $\mu\to e\gamma$ etc.,
wishing for the best, what else could happen?

Let us take one of the old tensions, $\sin2\beta$ vs $B \to \tau\nu$, as stressed by Descotes-Genon~\cite{Descotes-Genon} and Soni~\cite{Soni}, as an example.
If NP is on the $\sin2\beta$, or $CP$ phase of $B_d$-mixing side,
then perhaps $\sin\phi_s$, together with $K_L \to \pi^0\nu\nu$ could be interesting.
But $K_L \to \pi^0\nu\nu$ is still rather far away (and $K^+ \to \pi^+\nu\nu$
is a relatively blunt instrument).
On the $B\to \tau\nu$ side, on one hand, it is part of the
strength of $V_{ub}$ problem, where the implied $|V_{ub}|$ is
larger than even the inclusive value. Continued progress with lattice studies would help.
On the other hand, the recent findings of BaBar on $B\to D^{(*)}\tau\nu$
indicate that a charged Higgs $H^+$ of the type II 2HDM is insufficient
to describe the discrepancy from SM.
This rhymes well with some CKMfitter-type study done a while ago
regarding $B\to \tau\nu$, that if this is anomalous, it needs some NP
beyond $H^+$ of type II 2HDM~\cite{Descotes-Genon,Deschamps}.

Of course, Belle and BaBar should continue to scrutinize $B\to \tau\nu$,
where we await a major update from Belle~\cite{Hara}.
But, as reported in Koji Hara's talk~\cite{Hara},
even as it now stands, BaBar and Belle almost touch $B\to \mu\nu$,
which has the same $r_H$ enhancement factor~\cite{Hou93} as in Eq.~(1).
On closer scrutiny,
Belle's result~\cite{munuBelle} of ${\cal B}(B\to \mu\nu) < 1.7 \times 10^{-6}$ at 90\% C.L.
is based on analyzing 277 M $B\bar B$ pairs done in 2007,
while BaBar's limit~\cite{munuBaBar} of $1.0 \times 10^{-6}$ based on 468 M $B\bar B$ pairs
is somewhat better.
It seems to me that Belle should be able to improve the bound by at least a factor of two,
and I see no reason why the two experiments could not try to combine
the two datasets. If $B\to \tau\nu$ is genuinely enhanced, then
$B\to \mu\nu$ should be similarly enhanced, and this effort should provide a hint.

Which brings me to the illustration towards 2015+.
The Super B Factories are under construction, and SuperKEKB/Belle 2 should
be commissioned by 2015 (ahead of SuperB). With the luminosity gain of 40 to 50 times that of Belle,
it should be able to ``immediately" unravel the $B\to \mu\nu$ rate,
while $B\to \tau\nu$ will take a little longer.
But the $B\to \mu\nu$ analysis would have totally different systematics than $B\to \tau\nu$.
By then we would know whether there is a real hint for New Physics in such transitions,
and/or distinguish whether there may be some experimental bias.

Let us look forward to the super factories era.

\section{CONCLUSION: DON'T HOLD YOUR BREATH}

Let me give my conclusion as follows:
\begin{itemize}
\item There is no sign of Godot in Flavor Physics: LHCb eliminated all of them!
\item Some hope in $\sin\phi_s$, $B_s \to\mu\mu$, and $\mu\to e\gamma$.
\item Remaining tensions await 2015 or longer.
%
\item The Would-Be-Godot, the Higgs, may turn out a Pozzo.
\item Hold your breath until December,\\
 \phantom{ii}then exhale, and breath normally. \\
 Until 2015.
\end{itemize}

I will be most glad if these ``predictions" all become falsified;
this may well happen given the success of Altarelli and Teubert in 2011
(hence I wrote so intentionally).

Meanwhile, we do have Rio\footnote{
 Site for FPCP 2013.}
to look forward to!

\section{EPILOGUE: IS ``HIGGS" GODOT?}

At the big event at CERN held on July 4th, with simulcast
to the big ICHEP gathering in Melbourne half a globe away, there was the staggering announcement
that ``the Higgs" is bagged: both ATLAS and CMS see $\sim 5\sigma$ effect.
So, I am already heading the way of Altarelli and Teubert.
It goes without saying that this is a historic, landmark event.
Although it is quite orthogonal to the flavor world,
as commented on in Sec.~3, there would be repercussions.
For example, my beloved 4th generation is once again viewed to be in deep trouble.
Indeed it is. But as much as the world is round, flavor and electroweak are
almost orthogonal directions. Let us keep some separation between flavor and EWSB,
and see how things would develop towards FPCP 2013 (maybe FPCP 2016).

\bigskip 
\begin{acknowledgments}
My original intent on attending FPCP 2012 was just to pay
my deep respect to Prof. Zheng-guo Zhao, my former colleague at PSI.
It has certainly been an honor, therefore, to be asked to give the Theory Summary.
As Zheng-guo said in his closing remarks, the most impressive aspect of FPCP 2012
is the amount of open discussion, which I have indeed rarely seen.
This work is supported by
NSC 100-2745-M-002-002-ASP of the National Science Council of Taiwan,
and various NTU grants, including the President's office,
under the Ministry of Education ``Excellence" program.

\end{acknowledgments}

\bigskip 

\end{document}